\documentclass[aps,pre,twocolumn,showpacs,superscriptaddress]{revtex4}
\usepackage{amsfonts,amssymb,amsmath,times}
\usepackage{graphicx}
\usepackage{bm}
\usepackage{enumerate}
\usepackage{color}

\begin{document}

\title{Analyzing long-term correlated stochastic processes by means of
recurrence networks: Potentials and pitfalls}

\author{Yong Zou} 
%	\email[Email: ]{yzou@phy.ecnu.edu.cn}
 	\affiliation{Department of Physics, East China Normal University, 200062
 	Shanghai, China} 
 	\affiliation{State Key Laboratory of Theoretical Physics, Institute of
 	Theoretical Physics, Chinese Academy of Sciences, Beijing 100190, China}

\author{Reik V.~Donner}
%	\email[Email: ]{reik.donner@pik-potsdam.de}
	\affiliation{Potsdam Institute for Climate Impact Research, P.\,O.~Box
	60\,12\,03, 14412 Potsdam, Germany}

\author{J\"urgen Kurths}  
	\affiliation{Potsdam Institute for Climate Impact Research, P.\,O.~Box
	60\,12\,03, 14412 Potsdam, Germany}
 	\affiliation{Department of Physics, Humboldt University Berlin,
 	Newtonstra{\ss}e 15, 12489 Berlin, Germany} 
 	\affiliation{Institute for Complex Systems and Mathematical Biology,
 	University of Aberdeen, Aberdeen AB243UE, United Kingdom}
	\affiliation{Department of Control Theory, Nizhny Novgorod State University, Gagarin Avenue 23, 606950 Nizhny Novgorod, Russia}

\date{\today}

\begin{abstract}
Long-range correlated processes are ubiquitous, ranging from climate variables
to financial time series. One paradigmatic example for such processes is
fractional Brownian motion (fBm). In this work, we highlight the potentials and
conceptual as well as practical limitations when applying the recently proposed
recurrence network (RN) approach to fBm and related stochastic processes. In
particular, we demonstrate that the results of a previous application of RN
analysis to fBm (Liu \textit{et al.,} Phys. Rev. E \textbf{89}, 032814 (2014))
are mainly due to an inappropriate treatment disregarding the intrinsic
non-stationarity of such processes. Complementarily, we analyze some RN
properties of the closely related stationary fractional Gaussian noise (fGn)
processes and find that the resulting network properties are well-defined and
behave as one would expect from basic conceptual considerations. Our results
demonstrate that RN analysis can indeed provide meaningful results for
stationary stochastic processes, given a proper selection of its intrinsic
methodological parameters, whereas it is prone to fail to uniquely retrieve RN
properties for non-stationary stochastic processes like fBm.
\end{abstract}


\maketitle

\section{Introduction}

Many tools of nonlinear time series analysis are based on the theory of
(deterministic) dynamical systems~\cite{kantz1997,Sprott2003}, i.e., the time
evolution of the system under study is considered in some phase space spanned by
the relevant dynamical variables. Among others, the recurrence of previous
states in phase space~\cite{poincare1890} is a particular fundamental property
of dynamical systems with a finite phase space volume (e.g., attractors of a
dissipative system, Hamiltonian systems with a bound phase space, or even
stationary stochastic systems in finite time). The concept of recurrence implies
that the dynamics of a system returns to an arbitrarily small neighborhood of
any of its previously assumed states within a finite (but possibly large) amount
of time. For deterministic-chaotic systems, this is guaranteed by the invariance
of the set which forms the support of the attractor~\cite{Ott1993,Sprott2003}.

Recently, complex network representations have been proposed to characterize
statistical properties of the underlying system associated with its geometry in
phase space~\cite{Xu2008,Marwan2009,Donner2010NJP}. For this purpose, a proper
transformation from the set of state vectors in phase space to a complex network
representation is required. In this work, we focus on the recurrence network
(RN) approach, the vertices of the network are given by the individual
state vectors sampled from a given trajectory, whereas network connectivity is
established according to their mutual closeness in phase space (i.e., whether or
not their mutual distance is smaller than a pre-defined threshold
$\varepsilon$). Mathematically, given two state vectors $\mathbf{x}_i$ and
$\mathbf{x}_j$ (where $i$ and $j$ denote time indices associated with two
possibly different points $t_i$ and $t_j$ in time), the adjacency matrix
$A_{i,j}$ of the RN is defined as
\begin{equation}
A_{i,j}(\varepsilon) = \Theta(\varepsilon-\|\mathbf{x}_i-\mathbf{x}_j \|) - \delta_{i,j},
\label{eq_defrp}
\end{equation}
where $\Theta(\cdot)$ is the Heaviside ``function'', $\varepsilon$ is the
prescribed maximum distance, $\| \cdot\|$ a norm in phase space (e.g.,
Euclidean, Manhattan, or maximum norm), and $\delta_{i,j}$ is the Kronecker
delta. RN analysis originates from the recurrence plot
concept~\cite{marwan2007,Eckmann1987} and its basic assumption is, as the term
indicates, the unambiguous presence of recurrence behavior.

Stationarity is a condition required by most tools of both linear and nonlinear
time series analysis~\cite{kantz1997}, including the RN approach. A signal is
(strongly) stationary if all joint probabilities of finding the system at some
time in one state and at some later time in another state are independent of
time within the observation period. The minimal requirement for most approaches
is weak stationarity, that is, mean and variance of the underlying process are
constant and the auto-covariances depend only on the time lag.
  
In turn, many real-world processes are non-stationary. For instance, climate or
hydrological data often show seasonal variations. Economic and financial time
series typically exhibit (irregular) cycles of all orders of magnitude.
Non-stationary behaviors can be expressed in terms of trends, cycles, random
walks, or combinations of the latter three. Often, long-range dependence and
self-similarity are involved. One classical example of a class of such
non-stationary processes is fractional Brownian motion (fBm), which has
long-range temporal correlations as its defining property~\cite{Mandelbrot1968}.
Specifically, for an fBm process $\{X_t\}$, the variance scales as
$\sigma_{X_t}^2\propto t^{2H}$ (i.e., non-stationarity in variance). The long
range of the process is characterized by the Hurst exponent $H$ when positively
correlated (persistence) for $1/2 < H < 1$, while suppressed (anti-persistence)
for $0< H < 1/2$. $H=1/2$ corresponds to the classical Brownian motion.
  
Non-stationarity provides a great challenge to both linear and nonlinear time
series analysis, including complex network approaches to analyze time series
data. There are some methods that are specifically tailored to cope with
non-stationarity. Among others, detrended fluctuation analysis
(DFA)~\cite{Peng1994,Kantelhardt2001,Hu2002} and related techniques have been
widely used for estimating the Hurst exponent from non-stationary model data as
well as real-world applications from various fields~(e.g.,
\cite{Bunde2002,Kantelhardt2003}). In contrast, regarding the RN approach,
non-stationarity due to time-dependent system parameters can cause a systematic
loss of recurrences. Anyway, RNs have been recently proposed to characterize
fBm~\cite{Liu_PRE2014}. Notably, the results of the latter study have been
obtained only numerically and not explained theoretically so far. However, as we
will demonstrate in the course of this work, they have rather limited physical
interpretation. More generally, we will discuss how spurious results and
pitfalls of RN analysis may be produced when this method is inappropriately
applied to study fBm or other non-stationary stochastic processes, and that the
results of \cite{Liu_PRE2014} are mainly of such spurious nature.

This paper is organized as follows: In Section~\ref{sec:FBM}, we discuss the
construction of RNs from non-stationary fBm data. We demonstrate that it is not
possible to define generally applicable embedding parameters as required for a
systematic investigation of the potential effect of $H$ on the RN properties.
Specifically, we provide numerical evidence that the latter properties (for
given embedding parameters) depend explicitly on the system size, which
generally does not apply to stationary systems provided that the sample size is
large enough and sampling artifacts as well as transient behavior are avoided.
Subsequently, in Section~\ref{sec:FGN}, we turn to the RN properties of the
closely related fractional Gaussian noise (fGn), the incremental process
associated with fBm, which is stationary. For the latter, the dependence of the
network characteristics on $H$ is -- in contrast to fBm -- well-behaved.
However, the considered embedding dimension still plays an important role when
characterizing the RN structures. All results are summarized and further
discussed in Section~\ref{sec:conclusions}.

\section{RN analysis of fBm processes} \label{sec:FBM}

The application of RNs to the analysis of nonlinear time series implicitly
assumes the validity of two fundamental assumptions: (i) the intrinsic model
parameters and statistical characteristics of the system remain constant over
time and (ii) the system under study is sufficiently sampled (i.e., time
resolution and time series length are sufficient to approximate the system). The
first assumption is equivalent to the condition of stationarity, while the
second one mainly requires a proper coverage of phase space by a suitably
embedded time series. Both requirements are consequences of the fact that we
approach the system's dynamics by a single finite time series, which is common
to time series analysis problems. Note that there have been attempts to
characterize non-stationary systems by means of RN analysis using sliding
windows approaches, which have provided interesting results regarding the
presence of bifurcation or other qualitative changes in the dynamical
regime~\cite{Donges2011PNAS,Donges2011NPG}. However, these considerations have
been related to systems with supposed time-varying parameters rather than
non-stationary stochastic systems where the parameters are constant. Therefore,
this approach might not be helpful in the present context dealing with
non-stationary variance.

In the following, we will focus on two important algorithmic parameters of the
RN approach, embedding dimension and delay. The impact of other parameters such
as recurrence threshold $\varepsilon$, sampling rate, or even the selection of
variables in multi-dimensional systems has been extensively discussed
elsewhere~\cite{Donner2010PRE,strozzi2011} for deterministic systems, but not
yet for stochastic ones. For the sake of brevity, we present only a brief
corresponding discussion here. Specifically, since we consider discrete-time
univariate stochastic processes, only $\varepsilon$ is relevant, but can be
treated mostly alongside the theoretical considerations presented
in~\cite{Donges2012PRE}.

By means of conceptual considerations as well as numerical experiments, in the
remainder of this section, we will address the following three questions: (i)
Can we use embedding techniques for fBm (or, more generally, non-stationary
stochastic processes)? (ii) What are the intrinsic limitations of this approach?
(iii) Which implications do these limitations have for RN analysis?

\subsection{Time-delay embedding: Theoretical considerations} 

As the most prominent subject of recent studies involving RN
analysis~\cite{Marwan2009,Donner2010NJP,Donner2011IJBC,Zou2010Chaos,Zou2012Chaosa},
chaotic attractors exhibit some complex geometric structure in their respective
phase space, motivating the term ``strange attractors''. Typically, this
structure is associated with self-similar (fractal) characteristics. (Notably,
there are examples for strange non-chaotic attractors as
well~\cite{Grebogi1984,Feudel2006}.) Strange attractors emerge in deterministic
dynamical systems, and the resulting asymptotic set of state vectors approached
by the system forms some finite object in phase space. The dynamical properties
of the system and the geometric characteristics of the attractor are commonly
closely interrelated~\cite{Sprott2003,Donner2011EPJB}.

Taking this idea further, it is a natural approach to describe dynamical systems
of whatever kind by a geometric object in some appropriately defined phase
space. This is the basis of RN analysis, which takes the existence of such a
phase space (at least in an abstract sense) as a fundamental requirement. Given
this fact, RN analysis may be applied if the available data series provides
enough information to describe (or approximate) the geometric structure of the
sampled trajectory sufficiently. Specifically, transient dynamics has to be
excluded, data length and sampling frequency need to be appropriate, and the
data object in phase space needs to be dynamically invariant or at least bound
in phase space with stationary properties.

Given a scalar time series $\{x_i\}$ ($i=1,\dots,N$), in order to apply RN
analysis we first have to convert the data into state vectors in some
appropriately reconstructed phase space. A common method from dynamical systems
theory to define such a phase space is time-delay embedding~\cite{Takens1981}.
In fact, the concept of a phase space representation rather than a ``simple''
time or frequency domain approach is the hallmark of many methods of nonlinear
time series analysis, requiring embedding as the first step. Here, we define
$\mathbf{x}_i = (x_i, x_{i-\tau}, \cdots, x_{i-(m-1)\tau})$ to obtain an
$m$-dimensional time-delay embedding of $x_i$ with embedding delay $\tau$ for
obtaining state vectors in phase space~\cite{Takens1981}. It has been proven
that for deterministic dynamical systems, the thus reconstructed phase space is
topologically equivalent to the original space if $m > 2 D_F$, where $D_F$ is
the fractal dimension of the support of the invariant measure generated by the
dynamics in the true (but often at most partially observed) state space. Note
that $D_F$ can be much smaller than the dimension of the underlying original
(physical) phase space spanned by all relevant system variables.

From a practical perspective, when analyzing a scalar time series of whatever
origin, neither embedding dimension $m$ nor delay $\tau$ are known a priori. The
false nearest-neighbors (FNN) method~\cite{Kennel1992} was introduced to derive
a reasonable guess of how to choose $m$ based on studying whether or not
proximity relations between state vectors are lost when the embedding dimension
is successively increased. If a reasonable embedding dimension is found, all
dynamically relevant coordinates of the system are appropriately represented, so
that all proximity relationships are correct and not due to lower-dimensional
projection effects. In a similar spirit, the first root of the auto-correlation
function (ACF) of a time series often yields a good estimate for $\tau$. A more
refined method is to use time-delayed mutual information~\cite{Fraser1986}.

While the aforementioned approaches to determining $m$ and $\tau$ commonly work
well for data from deterministic dynamical systems, applying them to fBm leads
to severe conceptual problems:

On the one hand, we note that the concepts of a fractal dimension has two
aspects when being applied to a stochastic process instead of a deterministic
dynamical system. From the phase space perspective, the fractal dimension is
commonly defined as some scaling property described by a parameter that
converges to a fixed value as $N\to\infty$ and $m$ is sufficiently high. This
fact is used, for example, in the famous Grassberger-Procaccia algorithm for
estimating the correlation dimension $D_2$ of chaotic
attractors~\cite{grassberger83}. However, according to the latter viewpoint,
stochastic behavior is characterized by an absence of such convergence, formally
leading to $D_2=\infty$. Finite estimates of $D_2$ are spurious due to the
finite amount of data used. The latter result is reasonable since an infinite
amount of data (i.e., the innovations at each time step) are necessary to fully
describe the evolution of a stochastic process.

As an alternative perspective, the fractal dimension of a stochastic process is
often defined via the fractal dimension of its graph. For a one-dimensional
process, this graph is represented in the $(t,x)$-plane, and its dimension is
hence bound from above by $D_G=2$. Specifically, for fBm with a Hurst exponent
$H\in (0,1)$, it has been shown that $D_G=2-H$, taking the different scaling
behavior in association with the process' self-similarity into account
\cite{Mandelbrot1982,Gneiting2004}. However, the latter aspect is clearly
distinct from the notion of fractal dimensions used in the phase space context.
Thus, from a conceptual perspective, the embedding dimension should be chosen
infinitely large. In turn, finite $m$ will necessarily cause spurious results
since the full complexity of the system's (discrete) trajectory is not captured.

On the other hand, the embedding delay $\tau$ is not considered in the
mathematical embedding theorems for deterministic dynamical systems. Embeddings
with the same embedding dimension $m$ but different $\tau$ are topologically
equivalent in the mathematical sense~\cite{kantz1997}, but in reality a good
choice of $\tau$ facilitates further analysis. If $\tau$ is small compared to
the relevant internal time-scales of the system, successive elements of the
delay vectors are strongly correlated. This leads to the practical requirement
that the embedding delay should cover a much longer time interval than the
largest characteristic time-scale that is relevant for the dynamics of the
system. However, in fBm arbitrarily long time-scales are relevant due to the
self-similar nature of the process. This makes finding a feasible value of
$\tau$ a challenging (and, regarding formal optimality criteria, even
theoretically impossible) task.

In summary, we emphasize that in the case of non-stationary fBm, the fundamental
concepts of phase space reconstruction and low-dimensional dynamics do not apply
(not even approximately) anymore. Therefore, any attempt to applying RN analysis
to fBm directly necessarily yields results that hold only for the particular
embedding parameters chosen and the specific length of the given time
series~\cite{Liu_PRE2014}. We will demonstrate some numerical results
illustrating these points in more detail in the following.

\subsection{Numerical results} 

Estimating the ACF of a stationary time series at lag $\tau$ is straightforward
as long as $\tau$ is small compared to the total length of the time series, $N$.
For stationary stochastic processes, the functional form and rate of decay of
the ACF depends on the specific properties of the process. Specifically, for a
stationary long-range correlated process, the ACF decays like a power-law with
the characteristic exponent being directly related with
$H$~\cite{Witt_Geophy2013}.

\begin{figure}	
	\centering
	\includegraphics[width=\columnwidth]{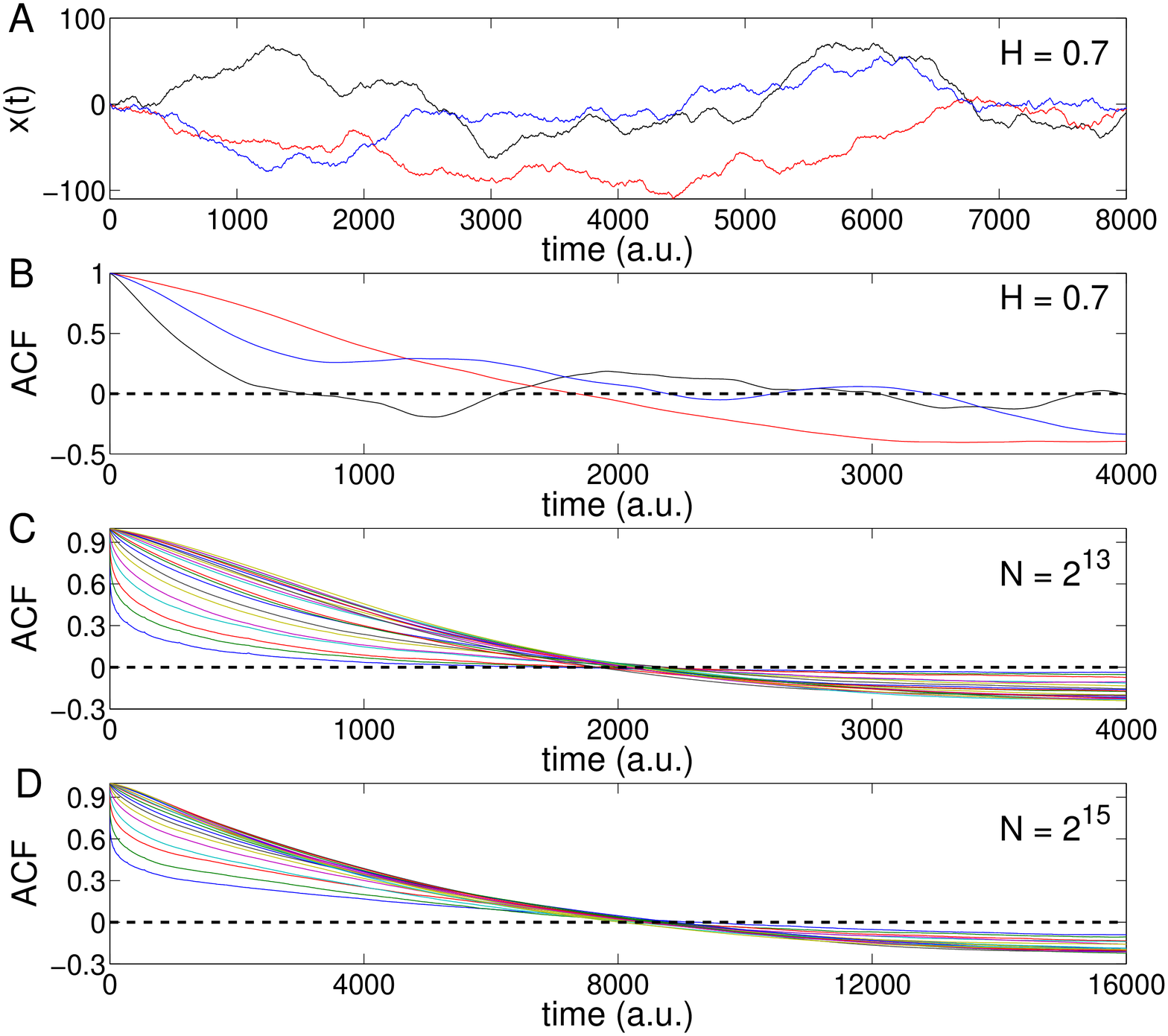}
\caption{(A) Three example trajectories of fBm with $H=0.7$ and (B) the corresponding
ACFs. (C) Average ACFs taken over 200 independent realizations of fBm with the
same Hurst exponent $H$. Different line colors correspond to different values of
$H$ from $0.05$ to $0.95$ in steps of $0.05$ (from bottom to top at small
$\tau$). In all cases, the time series length has been set to $N=2^{13}$. (D) As
(C) for $N=2^{15}$. \label{fig:fbm_acf}}
\end{figure}

In contrast to this, for the non-stationary fBm sample estimates of the ACF
decay extremely slowly beyond the ``normal'' behavior of stationary long-range
dependent processes, which can be seen clearly in Fig.~\ref{fig:fbm_acf} (in
fact, the concept of ACF is not appropriate for describing the serial dependence
structure of non-stationary processes). Specifically, we show three example
trajectories of fBms with $H=0.7$, $N=2^{13}$ and their corresponding na\"ive
ACF estimates. Due to the stochastic nature of the process, the de-correlation
time (which can be expressed as $\tau_{1/e}$ or $\tau_{0.1}$, i.e., the time
lags after which the estimated ACF has decayed to $1/e$ or $0.1$, respectively)
depends on the specific realization of the process (Fig.~\ref{fig:fbm_acf}B).
Even more, the corresponding ensemble spread does not exclusively originate from
the finite sample size, but is enhanced by the inherent non-stationarity of fBm.

Taking an ensemble average over a variety of independent realizations, we
numerically observe that the location of the first root of the estimated ACF
hardly depends at all on the Hurst exponent $H$, which is shown in
Fig.~\ref{fig:fbm_acf}C. However, as expected from theoretical study of fBm, it
appears to systematically increase as the length of time series is increased to
$N = 2^{15}$ (Fig.~\ref{fig:fbm_acf}D, note the different scales in
Figs.~\ref{fig:fbm_acf}C and D). More specifically, if we extend the length of
the realization by a factor of 4, the first root of the ACF estimate also shifts
to a four times larger lag.

Irrespective of the sample size $N$, the spectrum of the fBm process has a
significant amount of energy in frequencies that are not much larger than $1/N$
(i.e., in the low-frequency part). This explains why the first root of the ACF
estimate appears at larger time lags as $N$ is increased. Consequently, the
de-correlation time increases for longer time series. From the viewpoint of
time-delay embedding (given it is performed disregarding the conceptual concerns
detailed above), this hampers the proper choice of the embedding delay $\tau$.
In turn, the increasing persistence yields an increase in $\tau_{1/e}$ and
$\tau_{0.1}$ as well, as can be seen from the mutual offset of the different
lines in Fig.~\ref{fig:fbm_acf}C,D.

\begin{figure}
	\centering
	\includegraphics[width=\columnwidth]{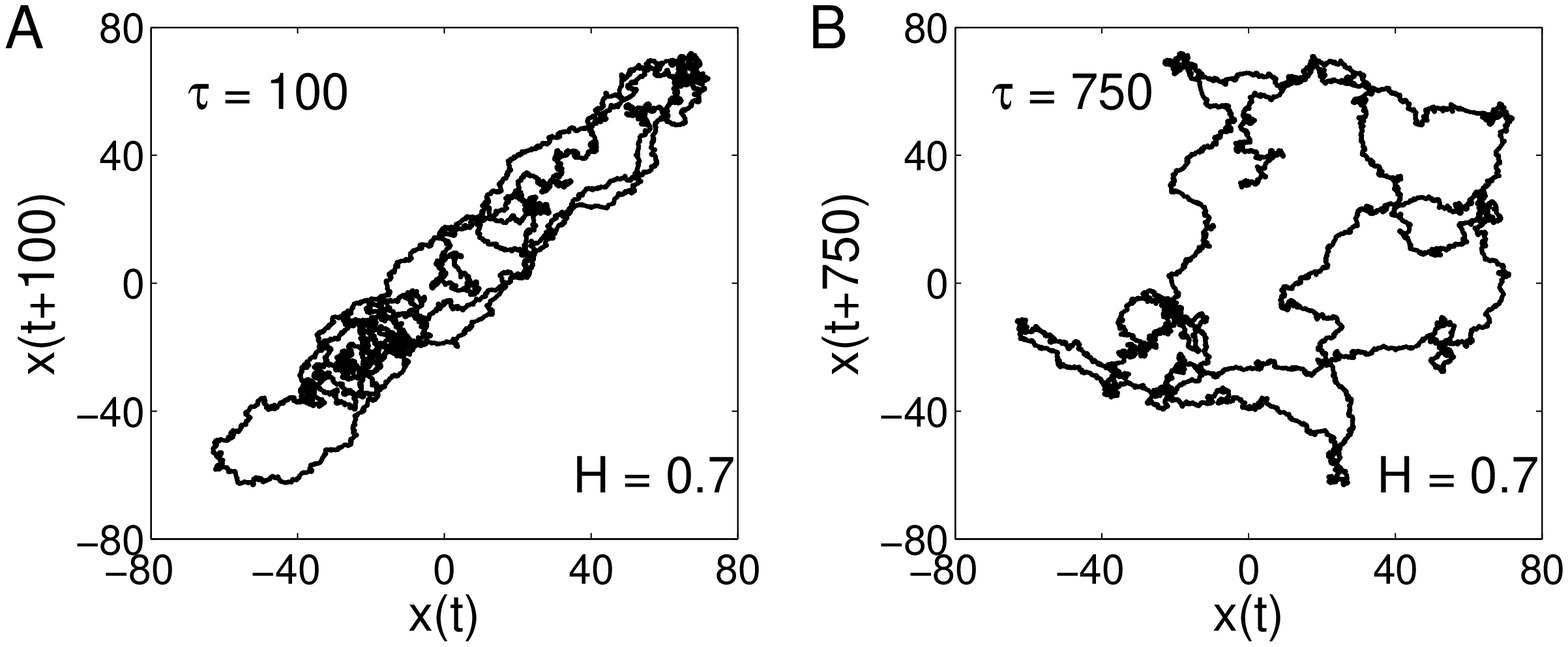}
\caption{Example trajectory of a fBm with $H=0.7$ in a two-dimensional reconstructed
phase space with embedding delays (A) $\tau=100$ and (B) $\tau=750$
($N=2^{13}$). \label{fig:fbm_phasespace}}
\end{figure} 

To further illustrate the practical consequences of the observed behavior of the
sample ACF when using embedding techniques, Fig.~\ref{fig:fbm_phasespace}
displays the same realization of a fBm embedded in a two-dimensional space with
different embedding delays $\tau$. Notably, the two embedding components are
highly correlated for small $\tau$ but less correlated for larger $\tau$,
leading to an entirely different geometric shape of the data object in the
reconstructed phase space. The same behavior will be necessarily observed in
higher embedding dimensions. As a consequence, a ``practical'' choice of the
embedding delay for fBm should be independent of $H$, but depend on $N$. The
numerical results presented above suggest $\tau\approx 2000$ for $N=2^{13}$ and
$\tau\approx 8000$ for $N=2^{15}$, possibly generalizing to $\tau\approx N/4$.
This is a rather large value, clearly far larger than those used by Liu
\textit{et~al.}~\cite{Liu_PRE2014} ($\tau\sim 10\dots 20$ for $N=2^{12}$).

\begin{figure}	
	\centering
	\includegraphics[width=\columnwidth]{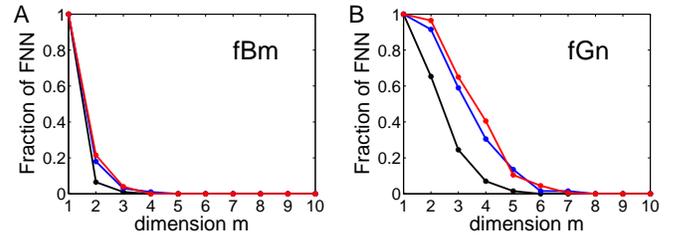}
\caption{Fraction of false nearest-neighbors (FNN) for (A) fBm and (B) fGn with $H=0.7$ and
different time series lengths $N$ (black: $N = 1000$, blue: $N = 21000$, and
red: $N = 31000$). \label{fig:fgn_fnn}}
\end{figure}

The determination of a reasonable embedding dimension $m$ is often achieved by
the FNN method~\cite{Kennel1992}. The criterion for the embedding dimension
being high enough is that the fraction of false nearest-neighbors is zero or at
least sufficiently small. Figure~\ref{fig:fgn_fnn}A displays our corresponding
numerical results for fBm for three different lengths, which consistently
suggest $m = 4$.

\subsection{Choice of the recurrence threshold}

An appropriate choice of the recurrence threshold $\varepsilon$ has attracted
great interest in the literature on
RNs~\cite{Donner2010PRE,Donner2011IJBC,Donges2012PRE,Donner2014Book,Eroglu2014}.
The most wide-spread procedure is fixing the resulting recurrence rate $\rho$
(i.e., the fraction of recurrences) and adjusting $\varepsilon$ accordingly. As
a rule-of-thumb, $\rho$ is often taken between about 0.01 and 0.05 for typical
RN sizes of a few thousand vertices~\cite{marwan2007,Donner2010PRE}, presenting
a trade-off between the necessity of avoiding a largely disconnected network
(too small $\epsilon$) and the interest in the geometric fine structure of the
system in its phase space, which is hidden when considering too large spatial
domains. The latter requirement has been more precisely formulated in
\cite{Donges2012PRE}, emphasizing on the empirically expected relationship for
the RN's average path length, $\mathcal{L}(\varepsilon)\sim\varepsilon^{-1}$,
which has been numerically confirmed \cite{Donner2010NJP,Donges2012PRE}.

Recently, \cite{Liu_PRE2014,Eroglu2014} suggested using the percolation
threshold of the random geometric graph constructed from the given distribution
of observed state vectors in phase space as a suitable lower bound to
$\varepsilon$. As shown by \cite{Donges2012PRE}, the scaling of the RN's average
path length breaks down if $\varepsilon$ falls below the limit for which the RN
decomposed into disjoint components, which is a necessary consequence of the
fact that the averaging involved in the calculation of $\mathcal{L}$ is commonly
considered only over pairs of vertices that are mutually
reachable~\cite{Donner2010NJP,Donges2012PRE}. However, when disregarding
shortest path-based RN characteristics, there is no reason why one should
restrict oneself to connected-networks, since other graph properties are hardly
affected by the presence of more than one component. In particular, requesting
the existence of a single component can lead to rather large $\varepsilon$ due
to the presence of outliers in the data~\cite{Donner2010PRE}, especially in case
of stochastic processes.

In this spirit, we recommend fixing $\rho$ at some reasonable value instead of
tuning $\varepsilon$ according to the percolation threshold. Notably, in this
case results obtained for different data sets still correspond to different
$\varepsilon$ when they originate from independent realizations of stochastic
processes. However, the problem of the dependence of some network measures on
the number of edges in the RN is relieved in this case. Note that for fBm, due
to the non-stationarity in variance the spread of state vectors in any
reconstructed phase space necessarily grows with the sample size $N$.

\section{RN analysis of fGn processes} \label{sec:FGN}

Based on our discussion presented in the previous section, we conclude that the
results recently presented in~\cite{Liu_PRE2014} hold only for the particular
choices of the algorithmic parameters (for instance, length of time series,
embeddings etc), showing limited physical interpretations. Moreover, using
non-stationary time series data necessarily produces unreliable and spurious
results.

One solution to the problem could be transforming the process in a way so that
it becomes stationary. In recent applications to non-stationary real-world time
series~\cite{Donges2011NPG,Donges2011PNAS}, the authors have removed
non-stationarities in the mean by removing averages taken within sliding windows
from the data. In the particular case of fBm, where non-stationary affects the
variance, the underlying stochastic process can be transformed into a stationary
one by a first-order difference filter, i.e., by considering its increments
$x_{i+1}- x_i$. The transformed series is commonly referred to as fractional
Gaussian noise (fGn) in analogy with the classical Brownian motion arising from
an aggregation of Gaussian innovations. Notably, fGn retains the long-range
correlations and Gaussian probability density function (PDF) from the underlying
fBm process. For illustration purposes, three independent realizations of fGn
with the same characteristic Hurst parameter $H=0.7$ are shown in
Fig.~\ref{fig:fgn_acf}A. Visual inspection clearly suggests the absence of
non-stationarity in both mean and variance.

\subsection{Embedding of fGn processes}

\begin{figure}	
	\centering
	\includegraphics[width=\columnwidth]{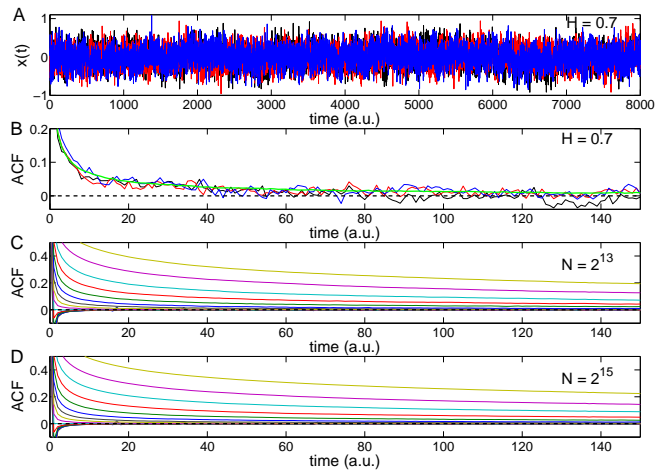}
\caption{As in Fig.~\ref{fig:fbm_acf} for fGn obtained by differencing the previous fBms.
In (B), the additional green line represents the averaged ACF over 200
independent realizations. \label{fig:fgn_acf}}
\end{figure}

Because of its stationarity, for fGn the estimated ACF shows a much faster decay
and less ensemble spread than for fBm (Fig.~\ref{fig:fgn_acf}B). Therefore,
disregarding the conceptual limitations of this approach when considering
stochastic processes, embedding parameters can be chosen more properly for fGn
than for fBm.  Concerning embedding delay $\tau$, one easily sees that $\tau =
1$ is a natural choice for $H<0.5$ according to the classical ACF criterion,
since the corresponding process is anti-persistent. Specifically, in this case
the ACF drops to negative value at lag one (as shown in
Fig.~\ref{fig:fgn_acf}C), i.e., subsequent values are negatively correlated --
the defining property of anti-persistence. In contrast, for $H > 0.5$ we use the
de-correlation time $\tau_{0.1}$ as an estimator for embedding delay $\tau$,
which increases with rising $H$ as one would expect since larger $H$ indicates a
longer temporal range of correlations.

As before, the embedding dimension $m$ is chosen via the FNN method. In
Fig.~\ref{fig:fgn_fnn}B, we show the fraction of false nearest-neighbors as $m$
is varied. Unlike for fBm, our results suggest that the optimal value $m$ rises
with an increasing length of the time series. In general, considerably higher
values of $m$ are suggested than for fBm, which matches the theoretical
expectations more closely. However, due to the finite sample size, we still find
a vanishing FNN rate at a finite embedding dimension, which is probably related
to a lack of proper neighbors when high dimensions are considered.

\subsection{Expected RN properties of stationary Gaussian processes}

Given a proper representation of the considered system by its phase space
reconstruction, the RN properties can be computed analytically from estimates of
the underlying $m$-dimensional state density
$p(\mathbf{x})$~\cite{Donges2012PRE}. In this spirit, an appropriate
representation requires that the sample size is sufficient to cover all relevant
parts of phase space, and that the sampling interval is reasonably chosen (i.e.,
to avoid sampling times co-prime with natural frequencies of continuous-time
systems). For fBm, the latter condition cannot be fulfilled due to the
non-stationarity of the process, whereas it is technically met for fGn
processes.

Making use of the analytical results of \cite{Donges2012PRE}, we expect that the
degree distribution $p(k)$ of the obtained RNs should be the same for any
stationary process with Gaussian PDF given the same embedding dimension $m$.
Specifically, this distribution has a complex shape \cite{Zou2012EPL} that is
independent of $H$ (note that we may fix the mean degree $\left<k\right>$ by
selecting a given $\rho=\left<k\right>/(N-1)$). In fact, this invariance is a
direct consequence of the fact that the geometry of the data in phase space is
not affected by $H$ when considering sufficiently de-correlated components, a
requirement that has not been met by \cite{Liu_PRE2014} in their recent
investigation of fBm as discussed above.

We emphasize again that the above considerations require a stationary Gaussian
process and an embedding for which all components are as close as possible to
being linearly independent. Otherwise, dependences between the components of the
embedding vector lead to a deformation of the data distribution in phase space
and, hence, possibly different geometric properties such as a too small
effective dimension (i.e., smaller than $m$).

\subsection{Transitivity properties}

In~\cite{Donner2011EPJB}, we have recently demonstrated that the RN
characteristics transitivity $\mathcal{T}$ and global clustering coefficient
$\mathcal{C}$ provide relevant information for characterizing the geometry of
the resulted RNs, which has been numerically supported for various
deterministic-chaotic systems. However, given the theory presented
in~\cite{Donges2012PRE}, the corresponding considerations can be extended to any
kind of process or, more generally, any kind of random geometric
graph~\cite{Dall_PRE_2002} with a given state density $p(\mathbf{x})$. Here, we
exemplify these considerations for the case of fGn and examine how the
transitivity properties of RNs arising from such stationary long-range
correlated stochastic processes depend on the characteristic Hurst exponent as
well as the underlying algorithmic parameters.
\begin{figure}	
	\centering
	\includegraphics[width=\columnwidth]{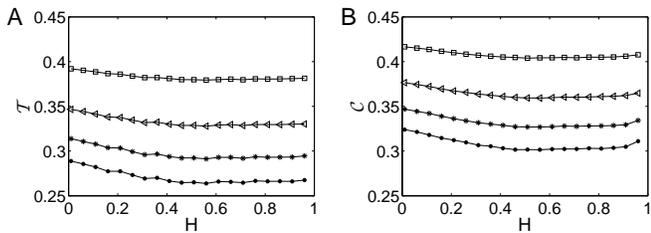}
\caption{Dependence of (A) RN transitivity $\mathcal{T}$ and (B) global clustering
coefficient $\mathcal{C}$ for fGn on the Hurst exponent $H$ for different
embedding dimensions ($m=3$: $\square$, $m=4$: $\triangleleft$, $m=5$: $\ast$,
$m=6$: $\bullet$), taken over 200 independent realizations and using a RN edge
density of $\rho=0.03$. The embedding delay has been kept at the same value for
all realizations with the same $H$ according to the de-correlation time
$\tau_{0.1}$. In all cases, $N=2^{12}$. \label{fig:fgn_trans_H}}
\end{figure}

\begin{figure}	
	\centering
	\includegraphics[width=\columnwidth]{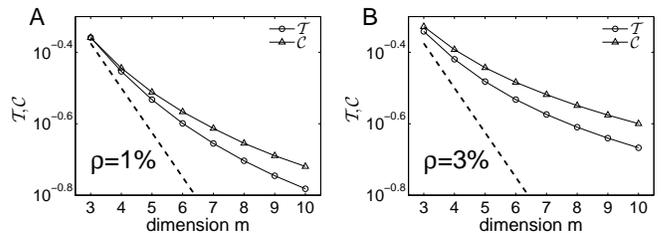}
\caption{Dependence of RN transitivity $\mathcal{T}$ and global clustering coefficient
$\mathcal{C}$ on the embedding dimension $m$ for fGn with $H=0.7$ (averages over
200 realizations) for two different values of $\rho$ ((A): $\rho=0.01$, (B):
$\rho=0.03$). The dashed line corresponds to the expected analytical values
$\mathcal{T}=(3/4)^m$ for $m$-dimensional Gaussian processes. In all cases,
$N=2^{12}$. \label{fig:fgn_trans_m}}
\end{figure}

For $H>0.5$, Fig.~\ref{fig:fgn_trans_H} shows that for a given embedding
dimension $m$, both transitivity and global clustering coefficient do not depend
on $H$. Following our above considerations, this is expected since the
$m$-dimensional Gaussian PDF of the process does not depend on $H$, and the
components are sufficiently de-correlated so that any marked geometric
deformation of the embedded data is avoided. Hence, we construct RNs from the
same PDF in all cases. Some minor deviation from the constant values can be
observed at $H$ close to 1, i.e., close to the non-stationary limit case
represented by $1/f$-noise, which might be due to numerical effects since the
corresponding processes are harder to simulate than such with moderate $H$.

For $H<0.5$, the behavior changes markedly: both $\mathcal{T}$ and $\mathcal{C}$
rise with decreasing Hurst exponent. The reason for this behavior is that
$\tau=1$ is the recommended, but still not ``optimal'' embedding delay for
anti-persistent processes. Specifically, the closer $H$ approaches 0, the
stronger is the anti-correlation at lag one. This means that with the same
embedding delay $\tau=1$, the smaller $H$ the stronger are the mutual
correlations between the different components of the embedding vector. As a
consequence, the state vectors do not form a homogeneous $m$-dimensional
Gaussian PDF with independent components in the reconstructed phase space, but
are stretched and squeezed along certain directions, so that the resulting
geometric structure appears significantly lower-dimensional than $m$. 

Given that $\mathcal{T}$ ($\mathcal{C}$) is related to a geometric notion of the
global (average local) dimension of the data~\cite{Donner2011EPJB}, a reduced
dimensionality of the data object results in a positive bias of both properties,
which is exactly what we observe here (Fig.~\ref{fig:fgn_trans_H}). Following
the latter considerations, it is also easy to explain why both $\mathcal{T}$ and
$\mathcal{C}$ systematically decrease with increasing embedding dimension $m$
(Figs.~\ref{fig:fgn_trans_H},~\ref{fig:fgn_trans_m}). Specifically, for a random
geometric graph in $m$ dimensions (computed with the maximum norm as also used
in this work), one can show analytically that
$\mathcal{T}=(3/4)^m$~\cite{Donner2011EPJB} (similar considerations apply to
$\mathcal{C}$~\cite{Donges2012PRE}). For a fixed sample size $N$, however, this
theoretical expectation is only met at low embedding dimensions $m$, whereas we
find a systematic upward bias of both $\mathcal{T}$ and $\mathcal{C}$ as $m$
increases (Fig.~\ref{fig:fgn_trans_m}). We explain the latter observation by the
finite sample size together with the problem that proximity relationships become
more ambiguous in higher dimensions when fixing a certain value of $\rho$.
Therefore, it can be expected that the bias should be systematically reduced
when using larger sample sizes $N$ together with smaller edge densities $\rho$
(for the latter effect, cf.~Fig.~\ref{fig:fgn_trans_m}A,B).

It would be straightforward to extend this kind of analysis to other network
measures, since the available analytical description of RNs allows for their
calculation as well~\cite{Donges2012PRE}. We leave a corresponding discussion as
a subject of future work.

\section{Conclusions}\label{sec:conclusions}

By a critical reassessment of previous work~\cite{Liu_PRE2014}, we have
identified several sources of errors when applying recurrence network analysis
(or, in a similar way, other concepts based on recurrences in phase space) to
long-range correlated stochastic processes. In summary, the main conclusions of
this analysis are as follows:

\begin{enumerate}[(i)]
\item RN analysis is based on phase space concepts originated in the theory of
deterministic dynamical systems. Therefore, its potential application to
stochastic processes requires special care.

\item The RN theory~\cite{Donges2012PRE,Donner2014Book} holds only for
stationary processes. A direct application of RN analysis to typical
non-stationary processes (in particular fBm) therefore has to fail, since the
PDF of the process in the considered phase space changes with time. Without
correcting for non-stationarity by a proper transformation of the series, the
obtained results are commonly spurious.

\item A major problem associated with non-stationary processes is that embedding
cannot be properly defined. In particular, the necessary selection of an
embedding delay is ambiguous since auto-correlation function and related
measures of serial dependences are not well-defined anymore.

\item For stationary stochastic processes, an embedding delay can be formally
estimated from the data. However, the problem of selecting an embedding
dimension remains, since stochastic processes are (in the viewpoint of dynamical
systems theory) infinite-dimensional. Hence, any low-dimensional embedding of a
stochastic process necessarily loses relevant information, which is a major
cause of spurious results.
\end{enumerate}

Despite the aforementioned conceptual problems and pitfalls resulting thereof,
RN can still be used for obtaining interesting information on stationary
stochastic processes. Drawing upon the interpretation of RNs as random geometric
graphs~\cite{Dall_PRE_2002} in some reconstructed phase space, the network
properties could in principle be computed solely from the multi-dimensional PDF
of the embedded process. Deviations from the expectations are related to
statistical dependencies between the different embedding components as well as
finite-sample and finite-scale effects. The latter are also relevant for
deterministic-chaotic processes, where in turn the underlying PDF can often not
be calculated or at least estimated with high accuracy. In this spirit, deriving
information based on stochastic processes can indeed help by providing
benchmarks for studies of deterministic dynamics.

In general, applying RN analysis to scalar measurements requires an appropriate
choice of embedding parameters. We do not claim that all choices made in this
work have been based on fully objective quantitative criteria. The concepts like
de-correlation time and false nearest-neighbors applied in this work rather
present heuristics capturing only some aspects relevant for obtaining a proper
phase space reconstruction. In this spirit, the results reported
in~\cite{Liu_PRE2014} are conceptually interesting but practically difficult to
interpret. For systematic applications, the choice of embedding parameters
depends on the particular process under consideration and should involve careful
statistical evaluation beyond visual inspection.

Finally, we emphasize that for non-stationary systems, embedding parameters
cannot be properly defined in general, so that any RN analysis (as well as other
time series analysis techniques) necessarily yields systematic errors. This
particularly applies to fBm and related processes arising from an integration of
stationary processes (e.g., fractional L\'evy motion, (F)ARIMA models, etc.). In
such cases, a proper transformation is required to remove the particular type of
non-stationarity from the data. This can be achieved by additive detrending,
phase adjustment (de-seasonalization), difference filtering (incrementation) or
other techniques, with the one mentioned last being the proper tool for the
particular case of fBm transforming the original process into stationary fGn.
Applying RN analysis to the latter indeed provides meaningful results. It should
be noted that this observation is consistent with some wide-spread conceptual
ideas beyond successful methodological alternatives for non-stationary time
series analysis such as DFA~\cite{Peng1994}, which commonly make use of
detrending and/or time series differentiation/aggregation. A more systematic
exploration of corresponding approaches in combination with recurrence-based
techniques is general, and RN analysis in particular, could be an interesting
subject of future work.

\subsection*{Acknowledgements.} 
YZ acknowledges financial support by the NNSF of China (Grant Nos. 11305062,
11135001), the Specialized Research Fund (SRF) for the Doctoral Program
(20130076120003), the SRF for ROCS, SEM, the Open Project Program of State Key
Laboratory of Theoretical Physics, Institute of Theoretical Physics, Chinese
Academy of Sciences, China (No.Y4KF151CJ1), and the German Academic Exchange
Service (DAAD). RVD has been funded by the Federal Ministry for Education and
Research (BMBF) via the Young Investigator's group CoSy-CC$^2$ (project no.
01LN1306A).

% \bibliography{ref_comment}

\end{document}